\documentclass[aps,showpacs,manuscript,12pt]{revtex4}
\usepackage{amssymb}
\usepackage{amsmath}
\usepackage{graphicx}

\begin{document}
\title{\textbf{IKT approach for quantum hydrodynamic equations$^{\S }$ }}
\author{M. Tessarotto$^{a,b}$, M. Ellero$^{c}$ and P. Nicolini$^{a,b}$}
\affiliation{\ $^{a}$Department
of Mathematics and Informatics, University of Trieste, Italy\\
$^{b}$Consortium for Magneto-fluid-dynamics\thanks{Web site:
http://cmfd.univ.trieste.it}, University of Trieste, Italy\\
$^{c}$Technical University of Munich, Germany}
\begin{abstract}
A striking feature of standard quantum mechanics is its analogy
with classical fluid dynamics. In particular it is well known the
Schr\"{o}dinger equation can be viewed as describing a classical
compressible and
non-viscous fluid, described by two (quantum) fluid fields $\left\{ \rho ,%
\mathbf{V}\right\} $, \ to be identified with the quantum
probability density and velocity field. \ This feature has
suggested the construction of a phase-space hidden-variable
description based on a suitable inverse kinetic theory (IKT;
Tessarotto et al., 2007). \ The discovery of this approach has
potentially important consequences since it permits to identify
the classical dynamical system which advances in time the quantum
fluid fields. This type of approach, however requires the
identification of additional fluid fields. These can be generally
identified with suitable directional fluid temperatures $T_{QM,i}$
(for $i=1,2,3$), to be related to the expectation values of
momentum fluctuations appearing in the Heisenberg inequalities.
Nevertheless the definition given previously for them (Tessarotto
et al., 2007) is non-unique. In this paper we intend to propose a
criterion, based on the validity of a constant H-theorem, which
provides an unique definition for the quantum temperatures.
\end{abstract}
\pacs{03.65.-w,05.20.Dd,05.20.-y}
\date{\today }
\maketitle



\section{Introduction}
The analogy with classical fluid dynamics has motivated in the
past efforts to formulate phase-space techniques based on various
statistical models of
quantum hydrodynamic equations. Following the pioneering work of Wigner \cite%
{Wigner1932,Bracken1999}, phase-space techniques typically require
that the quantum fluid fields or the wave-function itself be
represented in terms of, or associated with, appropriate
phase-space functions. These are usually identified with
quasi-probabilities \cite{Cahill} (for a review see Ref.
\cite{Gardiner-Zoller}, Chapter 4). This fact has motivated in the
past the search of alternative phase-space representations of the
quantum state. These works, although based on different
treatments, share the common view that the quantum state
corresponds to an underlying statistical description of some sort
(for a review of the phase-space route to the quantum hydrodynamic
equations see for example \cite{Wyatt}). A type of statistical
approach is provided by statistical models based on kinetic
theory. Examples
are provided by the kinetic models due to Kaniadakis \cite%
{Kaniadakis-2000-1,Kaniadakis-2003} and Goldstein and Pesci \cite%
{Pesci2005,Pesci2005b,Pesci2005c} , which are based on the
adoption of the Boltzmann kinetic equation to map quantum
hydrodynamic equations. However, also in this cases the relevant
wave equations are - at best - recovered only in an approximate
sense.

In this paper we intend to show that certain difficulties of
previous theories (in particular the issue of closure conditions
of quantum
hydrodynamic moment equations) can be overcome. For this purpose an \emph{%
inverse kinetic theory} is adopted for the Schr\"{o}dinger
equation, based on the introduction a suitable \emph{inverse
kinetic theory} (IKT) (Ref. A \cite{Piero}). By definition an IKT
\cite{Ellero2005,Ellero2006} must be able to yield identically,
via suitable moment equations, the complete system of fluid
equations. In particular, this can shown to satisfy the following
requirements:

\begin{enumerate}
\item \emph{completeness:} all fluid fields are expressed as moments of the
kinetic distribution function and all hydrodynamic equations can
be identified with suitable moment equations of IKE;

\item \emph{closure condition of moment equations:} there must exist a
subset of moments of IKE which form a complete system of
equations, to be identified with the prescribed set of quantum
hydrodynamic equations;

\item \emph{smoothness for the wave function}: the system wave-function is
assumed suitably smooth so that the solution of the kinetic
distribution function exists everywhere in a suitable extended
phase-space;

\item \emph{arbitrary initial and boundary conditions for the system
wave-function}: the initial conditions for the Schr\"{o}dinger
equations are set arbitrarily while Dirichlet boundary conditions
are considered for the system wave-function;

\item \emph{self-consistency}: the kinetic theory must hold for arbitrary
(and suitably smooth) initial conditions for the kinetic
distribution function. In other words, the initial kinetic
distribution function must remain arbitrary even if a suitable set
of its moments are prescribed at the initial time.

\item \emph{non-asymptotic IKT:} i.e., the correct hydrodynamic equations
must be recovered by the inverse kinetic theory independently of
any physical parameter characterizing the quantum hydrodynamic
equations.
\end{enumerate}

The formulation of a theory of this type involves also the
identification of an underlying classical dynamical system, in
terms of which all relevant observables and related expectation
values are advanced in time. This feature is potentially important
for numerical simulations both in computational fluid dynamics and
quantum mechanics, since the corresponding phase-space
trajectories, which determine uniquely the evolution of the fluid
fields, can thus be evaluated numerically. \ This permits the
development of Lagrangian particle simulation methods in fluid
dynamics which exhibit a low computational complexity
\cite{Tessarotto2004}.

From the mathematical viewpoint inverse kinetic theories can be
obtained, in principle, for arbitrary fluid equations, an example
being provided by the inverse kinetic theory recently developed
for the
incompressible Navier-Stokes equation by Ellero and Tessarotto\ \cite%
{Ellero2005,Ellero2006}. A basic prerequisite for the formulation
of an inverse kinetic theory of this type is, however, the proper
definition of the relevant \emph{quantum fluid fields }and their
identification with suitable momenta (to be denoted as
\emph{kinetic fluid fields}), which include the \emph{kinetic
temperature} as well as the related definition of
\emph{directional temperatures} (see below). However, the case of Schr\"{o}%
dinger equation is peculiar because, as is well-known, its related
fluid equations apparently depend only on two quantum fluid
fields, respectively, to be identified with the observables
quantum probability density and the quantum fluid velocity, while
the notions of quantum temperature and directional temperatures\
(to be identified with the corresponding kinetic moments) remain
in principle arbitrary. The problem is not merely of interest for
theoretical and mathematical research, but has potential relevance
also for the understanding of the fluid description of quantum
mechanics and of the underlying statistical models. Our motivation
is to exploit the analogy between classical and QM hydrodynamics
descriptions in order to prove that the quantum observables and
the fluid fields can formally be represented by means of a purely
classical statistical model. Although the mathematical equivalence
should not too hastily be regarded as implying physical
equivalence of the two formulations, this suggests that some
relevant classical reasonings can be transferred to SQM for the
construction of the inverse kinetic theory. This concerns, in
particular, the adoption of the so-called principle of entropy
maximization (PEM) for the Shannon information entropy
\cite{Jaynes1957,Guiasu1987} for the determination of the initial
condition for the kinetic distribution function. As a consequence,
one finds that a particular solution for the initial kinetic
distribution function corresponds to a generalized Maxwellian
kinetic distribution function, carrying prescribed number density,
flow velocity and directional temperatures. Nevertheless, also
non-Maxwellian kinetic distribution functions can be considered as
admissible initial conditions. Another interesting consequence of
the kinetic formulation is the formal description of SQM by means
of a classical dynamical system (to be denoted as
\emph{phase-space Schr\"{o}dinger dynamical system}). This can be
interpreted as a system of fictitious particles interacting with
each other only by means of an appropriate mean-field interaction
which depends on appropriate quantum fluid fields and moments of
the kinetic distribution function.\ Such a classical description
is realized by means of an appropriate \emph{kinetic correspondence principle%
}, whereby the physical observables, quantum fluid fields and
quantum hydrodynamic equations are respectively identified with
ordinary phase-space functions, kinetic moments of the kinetic
distribution function and moment equations obtained from IKE. \ In
principle infinite solutions to this problem exist due to the
non-uniqueness of the definition of the kinetic directional
temperatures. Nevertheless, \ here we intend to show that the
adoption of the principle of constant Shannon entropy for isolated
quantum systems permits to determine them uniquely.

\section{Hydrodynamic description of NRQM}

The fluid description of non-relativistic quantum mechanics (NRQM)
is
well-known \cite{Madelung1928}. It is based on the property of the Schr\"{o}%
dinger equation to be equivalent to a complete set of fluid
equations, i.e., respectively the continuity and Euler equations
\begin{eqnarray}
&&\left. \frac{D\rho }{Dt}+\rho \nabla \cdot \mathbf{V}=0,\right. \\
&&\frac{D}{Dt}\mathbf{V}(\mathbf{r},t)=\frac{1}{m}\mathbf{F}\equiv -\frac{1}{%
m}\nabla U_{QM}.
\end{eqnarray}%
which are assumed to hold in a domain \ $\Omega \subseteq
\mathbf{R}
^{3}.$ These are known as \emph{quantum hydrodynamic equations
(QHE)}.\ For the sake of clarity, let us recall the basic
definitions and the mathematical formulation of the problem. Let
us consider, \ for definiteness the equation for one spinless
scalar particle (boson) described by a single scalar wavefunction
$\psi (\mathbf{r},t)=\sqrt{\rho }e^{i\frac{S}{\hbar }},$
where $\rho =\left\vert \psi (\mathbf{r},t)\right\vert ^{2}$ and $S(\mathbf{r%
},t)$\ are two smooth real function denoting respectively the
quantum probability density and the \emph{quantum phase-function}
(also denoted as Hamilton-Madelung principal function).
Furthermore,
\begin{equation}
\mathbf{V}(\mathbf{r},t)=\frac{1}{m}\nabla S(\mathbf{r},t),
\end{equation}%
is the quantum velocity field, while $U_{QM}$ is the so-called
quantum potential
\begin{equation}
U_{QM}=-\frac{\hbar ^{2}}{2}\left( \frac{1}{2}\nabla ^{2}\ln f+\frac{1}{4}%
\left\vert \nabla \ln f\right\vert ^{2}\right) +U,
\end{equation}%
with $U$ a suitably smooth real function denoting an appropriate
interaction potential. \ For well-posedness, appropriate initial
and boundary conditions
must be imposed on the fluid fields $\left\{ \rho (\mathbf{r},t),V(\mathbf{r}%
,t)\right\} .$ In particular, $\left\{ \rho (\mathbf{r},t),V(\mathbf{r}%
,t)\right\} $ are both assumed continuous in $\overline{\Omega
}\times I$ and at least $C^{(2,1)}(\Omega \times I).$ The set
fluid equations for the quantum fluid fields $\left\{
f,\mathbf{V}\right\} $ provide a complete description of quantum
systems. \ This means, in particular, that \emph{no other}
independent observable (or dynamical variable) is required to
describe the quantum state. However, it is useful to introduce the
concept of quantum directional temperatures. The definitions can
be obtained invoking the Heisenberg theorem, which requires (for
$i=1,2,3$)
\begin{equation*}
\left\langle \left( \Delta r_{i}\right) ^{2}\right\rangle
\left\langle \left( \Delta p_{i}\right) ^{2}\right\rangle \geq
\frac{\hbar ^{2}}{4},
\end{equation*}%
where $\overline{\Delta }r_{i}=\left\langle \left( \Delta
r_{i}\right) ^{2}\right\rangle ^{1/2},\overline{\Delta
}p_{i}=\left\langle \left( \Delta
p_{i}\right) ^{2}\right\rangle ^{1/2}$ \ (for $i=1,2,3$) are the \emph{%
quantum standard deviations }for\emph{\ }position and (quantum)
linear momentum and\ $\left\langle \left( \Delta r_{i}\right)
^{2}\right\rangle ,\left\langle \left( \Delta p_{i}\right)
^{2}\right\rangle $ the corresponding \emph{average quadratic
quantum fluctuations. }It follows that
\begin{equation}
\left\langle \left( \Delta p_{j}\right) ^{2}\right\rangle
=\left\langle \left( \Delta ^{(1)}p_{i}\right) ^{2}\right\rangle
+\left\langle \left( \Delta ^{(2)}p_{i}\right) ^{2}\right\rangle ,
\end{equation}
where
\begin{equation}
\left\langle \left( \Delta ^{(1)}p_{i}\right) ^{2}\right\rangle
=\frac{\hbar ^{2}}{4}\int\limits_{\Omega }d\mathbf{r}f\left(
\partial _{j}\ln f\right) ^{2},
\end{equation}%
\begin{equation}
\left\langle \left( \Delta ^{(2)}p_{i}\right) ^{2}\right\rangle
\equiv \left\langle \left( \partial _{j}S\right) ^{2}\right\rangle
-\left\langle
\partial _{j}S\right\rangle ^{2}.
\end{equation}
Hence the first term [on the r.h.s. of Eq.()] is by definition
strictly positive. By analogy with classical statistical\
mechanics, it is therefore
natural to introduce the notion of \emph{\ quantum directional temperatures }%
$T_{i}(t)$ (for $i=1,2,3$)
\begin{eqnarray}
T_{i}(t) &\equiv &T_{0}(t)+T_{QM,i}(t),  \label{Ti} \\
T_{QM,i}(t) &=&\frac{1}{m}\left\langle \left( \Delta
^{(1)}p_{i}\right) ^{2}\right\rangle ,  \label{Tii}
\end{eqnarray}%
to be viewed as a \emph{constitutive equation for} $T_{i}$. By
definition each $T_{i}$ results strictly positive, while
$T_{0}(t)$ is a strictly positive smooth real function of time to
be suitably defined.

\section{IKT for QHE}

Let us now set the problem of searching an inverse kinetic theory
for the Schr\"{o}dinger equation, i.e., a kinetic theory yielding
exactly, by means of suitable moment equations, the quantum
hydrodynamic equations. The theory must hold for arbitrary (and
suitably smooth) initial and boundary conditions both for the
wavefunction and the kinetic probability density. The form of the
quantum fluid equations suggests that they can be obtained as
moment equations of a continuous inverse kinetic theory, analogous
to that developed recently for the incompressible Navier-Stokes
equation. In the sequel we consider, without loss of generality,
the case of one-body quantum systems; the theory here developed is
applicable, in fact, with minor changes also for arbitrary
$N$-body systems with $N>1$ particles. For
definiteness, let us assume that the quantum fluid fields $\left\{ f(\mathbf{%
\mathbf{r}},t),\mathbf{V,}T_{i}\right\}$, for $ i=1,2,3,$ are
respectively solutions of the QHE indicated above and of the
constitutive equation (\ref{Ti}). To restrict the class of
possible kinetic models,
following the approach of Ref. A, let us introduce\ a probability density $g(%
\mathbf{x},t),$ with $\mathbf{x=}\left( \mathbf{r,v}\right) ,$
defined in
the phase space $\overline{\Gamma }=\overline{\Omega }\times U$ (where $%
U\equiv
\mathbf{R}
^{3N})$ and assume that in the open set $\Gamma =\Omega \times U$
the probability density $g(\mathbf{\mathbf{r,v},}t)$ satisfies a
Vlasov-type kinetic equation of the form (\emph{Assumption \#2})
\begin{equation}
\emph{L}g(\mathbf{x,}t)=0  \label{inverse kinetic equation -A}
\end{equation}%
(\emph{inverse kinetic equation}), where \emph{L} is the Vlasov
streaming operator
\begin{equation}
L=\frac{\partial }{\partial t}+\frac{\partial }{\partial
\mathbf{x}}\cdot \left( \mathbf{X}\right) \equiv
\frac{d}{dt}+\frac{\partial }{\partial \mathbf{v}}\cdot \left(
\frac{\mathbf{K}}{m}\right) , \label{streaming operator}
\end{equation}%
where $\mathbf{X}$ now indicates the vector field
\begin{equation}
\mathbf{X=}\left\{ \mathbf{v,}\frac{1}{m}\mathbf{K}\right\} ,
\label{X vector field}
\end{equation}%
$\mathbf{x=(r,v)}$ and $\mathbf{K}(\mathbf{x,}t),$ to be denoted as \emph{%
mean field force,} is a suitably smooth real vector field. The
functional
class $\left\{ g(\mathbf{x},t)\right\} $ and the vector field $\mathbf{K}(%
\mathbf{x,}t)$ are determined in such a way that:

\begin{enumerate}
\item $g(\mathbf{x},t)$ is non-negative and continuous in $\overline{\Gamma }%
\times I,$ in particular, is strictly positive and of class $%
C^{(2+k,1+h)}(\Gamma \times I),$ with $h,k\geq 1;$

\item $\forall \left( \mathbf{r,}t\right) \in \overline{\Omega }\times I,$ $%
g(\mathbf{x},t)$ admits the kinetic moments $M_{X}\left[ g\right]
\equiv
\int_{U}d\mathbf{vX}g,$ with $\mathbf{X}(\mathbf{\mathbf{r,v},}t)=1,\mathbf{%
v,}u_{i}^{2}$ (for $i=1,2,3$)$,\mathbf{uu,u}u^{2},\ln g,$ where
$\mathbf{v,}$ $\mathbf{u=v-V}$ are respectively the kinetic and
the relative kinetic velocities and $u_{i}=v_{i}-V_{i}$ (for
$i=1,2,3$) are the orthogonal Cartesian components of $\mathbf{u}$
defined with respect to an arbitrary inertial reference frame;

\item The kinetic moments $\mathbf{X}(\mathbf{\mathbf{r,v},}t)=1,\mathbf{v,}%
u_{i}^{2}$ (for $i=1,2,3$) satisfy a suitable set of constraint
equations, to be denoted as \emph{kinetic correspondence
principle,} namely the following equations are assumed to hold
identically for $i=1,2,3$
\begin{eqnarray}
&&f=M_{1}\left[ g\right] \equiv \int_{U}d\mathbf{v}g(\mathbf{\mathbf{r,v},}%
t),  \label{moment - 1A} \\
&&M_{2}\left[ g\right] \equiv \frac{1}{f(\mathbf{r},t)}\int_{U}d\mathbf{vv}g(%
\mathbf{\mathbf{r,v},}t)=  \label{moment -2A} \\
&=&\mathbf{V}(\mathbf{\mathbf{\mathbf{r}}},t\mathbf{),}  \notag
\end{eqnarray}%
\begin{eqnarray}
&&\left. M_{3i}\left[ g\right] \equiv \right.  \label{moment -3A} \\
&\equiv &\frac{1}{f(\mathbf{r},t)}\int_{U}d\mathbf{v}mu_{i}^{2}g(\mathbf{%
\mathbf{r,v},}t)=T_{i}(t)>0.  \notag
\end{eqnarray}

\item the kinetic equation admits local Maxwellian kinetic equilibria for
arbitrary kinetic moments and quantum fluid fields which satisfy
the kinetic correspondence principle (\ref{moment -
1A}),(\ref{moment -2A}),(\ref{moment -3A});

\item The vector field $\mathbf{X(x,}t)$ is prescribed in such a way that it
depends, besides on $\mathbf{x},$\ only on the fluid fields and
suitable differential operators acting on them.
\end{enumerate}

In Ref. A it was proven that, provided the directional quantum temperatures $%
T_{i}(t)$ are considered as prescribed , the vector field $\mathbf{K}(%
\mathbf{\mathbf{r,v},}t)$ which satisfies the assumptions
indicated above
can be uniquely constructed and reads \cite{Piero}%
\begin{eqnarray}
&&\left. \mathbf{K}(g)=m\mathbf{u}\cdot \frac{\partial }{\partial \mathbf{r}}%
\mathbf{V+}\right.  \notag \\
&&+\frac{m}{2}u_{i}\mathbf{e}_{i}\left[ \frac{\partial }{\partial
t}\ln
T_{i}+\frac{1}{\rho T_{i}}\mathbf{\nabla \cdot Q}\right] + \\
&&+\mathbf{F}(\mathbf{r,}t)+\frac{m}{\rho }\mathbf{\nabla \cdot }\underline{%
\underline{\Pi }},  \notag
\end{eqnarray}%
where $\boldsymbol{Q=}\int d\mathbf{v}\frac{1}{3}\mathbf{u}u^{2}g$ and $%
\underline{\underline{\Pi }}=\int d\mathbf{vuu}g.$

Nevertheless, the complete specification of the IKT requires the
unique definition of the observables\ $T_{i}(t).$ \ In fact, the
definition of the kinetic directional temperatures remains in
principle arbitrary since they do not enter explicitly the quantum
hydrodynamic equations. \ This means that there actually exist
infinite equivalent realizations of the IKT for the
Schr\"{o}dinger equation. This poses, therefore, the issue whether
uniqueness can be established based on first principles.

\section{Constant H-theorem for the Shannon kinetic entropy}

Let us now introduce the Shannon entropy for the kinetic
probability density $g(x,t)$\emph{\ }
\begin{equation}
S(g)=-\int\limits_{\Gamma }d\mathbf{x}g\ln g,
\end{equation}%
(\emph{Shannon kinetic entropy). }We intend to prove that provided
the function\emph{\ }$T_{0}(t)$ is suitably prescribed $S(g)$ can
be imposed to be constant in time. \ For this purpose let us
evaluate the entropy
production:%
\begin{equation*}
\frac{\partial }{\partial t}S(g)=\int\limits_{\Gamma }d\mathbf{x}g\frac{%
\partial }{\partial \mathbf{v}}\cdot \frac{1}{m}\mathbf{K}(f).
\end{equation*}%
Then the following result holds:\bigskip

\textbf{Theorem - Constant H-theorem}

\emph{Let us assume that the quantum system fulfills the following
assumptions:} \emph{(1) the initial-boundary-value problem associated to the Schr\"{o}%
dinger equation admits a suitably smooth global solution (i.e.,
which exists and is unique for all }$t\in
\mathbf{R}
$\emph{);} \emph{(2) the following integrals exist all }$t\in
\mathbf{R}
$\emph{:}%
\begin{equation*}
I_{1}=\int\limits_{\Omega }d\mathbf{r}g\nabla \cdot \mathbf{V}
\end{equation*}%
\begin{equation*}
I_{2}=\frac{1}{2}\int\limits_{\Omega }d\mathbf{r\nabla \cdot Q}%
\sum\limits_{i=1,2,3}\frac{1}{T_{i}}
\end{equation*} \emph{(3) that the directional quantum temperatures, are defined by
equations (\ref{Ti}) for all }$t\in I;$ \emph{(4) we impose the
initial condition} $T_{0}(t_{o})=T_{o},$ \emph{where
}$T_{o}>0$\emph{\ is an appropriate constant}$.$

\emph{It follows that\ }$T_{0}(t),$\emph{\ is uniquely determined for all }$%
t\in I$ \emph{and for} $t\geq t_{o}$\emph{\ by imposing that there
results identically:}

\begin{equation}
\frac{\partial }{\partial t}S(g)=0.
\end{equation}%
PROOF

The proof follows by explicit evaluation of the entropy production
rate,
which delivers\emph{\ }%
\begin{eqnarray}
&&\left. \frac{\partial }{\partial t}S(g)=\int\limits_{\Omega }d\mathbf{r}%
g\nabla \cdot \mathbf{V+}\right. \\
&&+\frac{1}{2}\left[
\sum\limits_{i=1,2,3}\frac{1}{T_{i}}\frac{\partial
T_{QM,i}}{\partial t}+\int\limits_{\Omega }d\mathbf{r\nabla \cdot Q}%
\sum\limits_{i=1,2,3}\frac{1}{T_{i}}\right] + \\
&&\left. +\frac{\partial T_{0}}{\partial t}\frac{1}{2}\sum\limits_{i=1,2,3}%
\frac{1}{T_{i}}\equiv 0.\right.
\end{eqnarray}%
This is an ode for $T_{0}(t)$ which under suitable smoothness
assumptions admits an unique solution.

\section{Conclusions}

Motivated by the analogy between hydrodynamic description of SQM
and classical fluid dynamics an inverse kinetic theory has been
developed for the quantum hydrodynamic equations. We have shown
that, although in principle infinite solutions to this problem
exist (in particular due to the indeterminacy in the kinetic
directional temperatures), the inverse kinetic theory can be
uniquely determined, provided appropriate hypotheses are
introduced. This includes the validity of a constant H-theorem for
the Shannon kinetic entropy. \ This requirement, on the other
hand, is consistent with the formulation of an IKT, both for
classical and quantum fluids. The conclusion here presented is
potentially relevant for the fluid description of quantum
mechanics and a deeper understanding of the underlying statistical
description. In particular, we stress that the concept of Shannon
kinetic entropy, here adopted, which is completely analogous to
the one adopted for classical systems, is potentially relevant
also for the investigation of irreversible phenomena in
non-isolated quantum systems.

\section*{Acknowledgments}
Work developed (M.T.) in the framework of the MIUR (Italian
Ministry of University and Research) PRIN Research Program
``Modelli della teoria cinetica matematica nello studio dei
sistemi complessi nelle scienze applicate'' and the European COST
action P17 (M.T). The partial of the GNFM (National Group of
Mathematical Physics) of INDAM (National Institute of Advanced
Mathematics, Italy) (M.T. and P.N.) and of the Deutsche
Forschungsgemeinschaft via the project EL503/1-1 (M.E.) is
acknowledged.

\section*{Notice}
$^{\S }$ contributed paper at RGD26 (Kyoto, Japan, July 2008).
\newpage




\newpage
\begin{thebibliography}{BIBTEX}
\bibitem{Wigner1932} E.P. Wigner, Phys.Rev. \textbf{40}, 749 (1932).

\bibitem{Bracken1999} A.J. Bracken, H.-D. Doebner and J.G. Wood,
Phys.Rev.Lett \textbf{83}, 3758 (1999).

\bibitem{Cahill} K.E. Cahill and R.J. Glauber, Phys. Rev. \textbf{177,} 1882
(1969).

\bibitem{Gardiner-Zoller} C.W. Gardiner and P. Zoller, \textit{Quantum
Noise, }3rd Edition (Springer Berlin, Heidelberg New York, 2004).

\bibitem{Wyatt} R. Wyatt, \textit{Quantum Dynamics with Trajectories},
(Springer, Berlin, Heidelberg New York, 2005.

\bibitem{Kaniadakis-2000-1} G. Kaniadakis, Physica A \textbf{307}, 172
(2002).

\bibitem{Kaniadakis-2003} G. Kaniadakis, Found. Phys. Lett. \textbf{16}(2),
99 (2003).

\bibitem{Pesci2005} A. Pesci and R. Goldstein, Nonlinearity \textbf{18}, 211
(2005).

\bibitem{Pesci2005b} A. Pesci and R. Goldstein, Nonlinearity \textbf{18},
227 (2005).

\bibitem{Pesci2005c} A. Pesci and R. Goldstein, Nonlinearity \textbf{18},
1295 (2005).

\bibitem{Piero} M. Tessarotto, M. Ellero and P. Nicolini, Phys.Rev. A\textbf{%
75}, 060691 (2007); arXiv:quantum-ph/060691.

\bibitem{Ellero2005} M. Ellero and M. Tessarotto, Physica A \textbf{355},
233 (2005).

\bibitem{Ellero2006} M. Tessarotto and M. Ellero, Physica A \textbf{373},
142 (2007); arXiv.org/physics/0602140 (2007).

\bibitem{Tessarotto2004} M. Tessarotto and M. Ellero, Proc. 24th
RGD, Bari, Italy (July 2004), Ed. M. Capitelli, AIP Conf.
Proceedings \textbf{762}, 108 (2005).

\bibitem{Jaynes1957} E.T. Jaynes, Phys. Rev. \textbf{106}, 620 (1957).

\bibitem{Guiasu1987} S. Guiasu, Phys.Rev.A \textbf{36}, 1971 (1987).

\bibitem{Madelung1928} E. Madelung, Zeit. F. Phys. \textbf{40}, 322 (1927).
\end{thebibliography}
\end{document}